# Quasi van der Waals Epitaxy of Rhombohedral-stacked Bilayer WSe$_2$ on GaP(111) Heterostructure

Aymen Mahmoudi[1], Meryem Bouaziz[1], Niels Chapuis[2], Geoffroy Kremer[1], Julien Chaste[1], Davide Romanin[1], Marco Pala[1], François Bertran[3], Patrick Le Fèvre[3], Iann C. Gerber[4], Gilles Patriarche[1], Fabrice Oehler[1], Xavier Wallart[2], Abdelkarim Ouerghi[1*]

[1]Université Paris-Saclay, CNRS, Centre de Nanosciences et de Nanotechnologies, 91120, Palaiseau, Paris, France
[2] Univ. Lille, CNRS, Centrale Lille, JUNIA ISEN, Univ. Polytechnique Hauts de France, UMR 8520-IEMN F59000 Lille France
[3]Synchrotron SOLEIL, L'Orme des Merisiers, Départementale 128, 91190 Saint-Aubin, France
[4]Université de Toulouse, INSA-CNRS-UPS, LPCNO, 135 Avenue de Rangueil, 31077 Toulouse, France

## Abstract

The growth of bilayers of two-dimensional (2D) materials on conventional 3D semiconductors results in 2D/3D hybrid heterostructures, which can provide additional advantages over more established 3D semiconductors while retaining some specificities of 2D materials. Understanding and exploiting these phenomena hinge on knowing the electronic properties and the hybridization of these structures. Here, we demonstrate that rhombohedral-stacked bilayer (AB stacking) can be obtained by molecular beam epitaxy growth of tungsten diselenide (WSe$_2$) on gallium phosphide (GaP) substrate. We confirm the presence of 3R-stacking of the WSe$_2$ bilayer structure using scanning transmission electron microscopy (STEM) and micro-Raman spectroscopy. Also, we report high-resolution angle-resolved photoemission spectroscopy (ARPES) on our rhombohedral-stacked WSe$_2$ bilayer grown on GaP(111)B substrate. Our ARPES measurements confirm the expected valence band structure of WSe$_2$ with the band maximum located at the Γ point of the Brillouin zone. The epitaxial growth of WSe$_2$/GaP(111)B heterostructures paves the way for further studies of the fundamental properties of these complex materials, as well as prospects for their implementation in devices to exploit their promising electronic and optical properties.

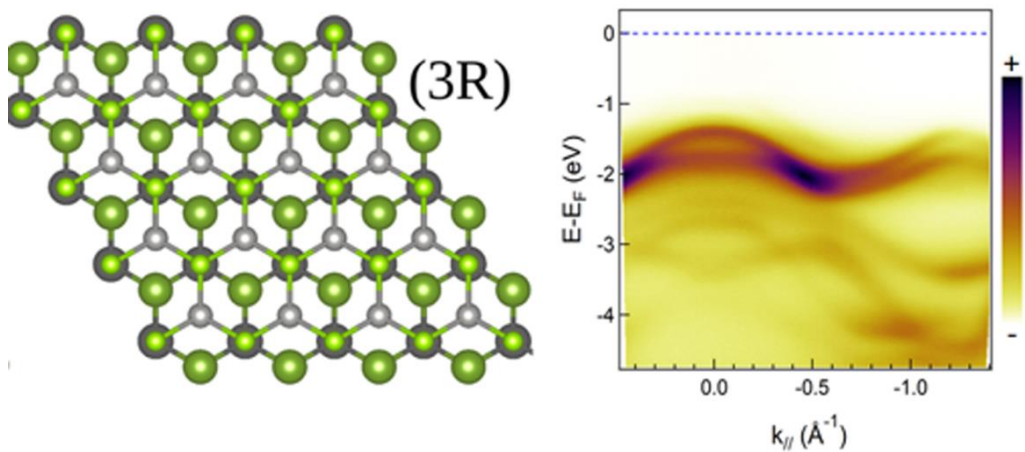

Keywords: 2D materials, MBE, 2D on 3D heterostructure, electronic band structure, Rhombohedral-stacked Bilayer WSe$_2$, density functional theory calculations

**Corresponding Author:** abdelkarim.ouerghi@c2n.upsaclay.fr

Emerging two-dimensional (2D) layered semiconductors have shown considerable potential for the design of next-generation integrated electronic and optoelectronic systems, due to their peculiar physical and structural properties [1]. In particular, diverse 2D layered semiconductors can be easily combined to form a variety of vertical van der Waals (vdW) heterostructures, with atomically sharp interfaces and tunable band alignment [2]. This helps for the fundamental investigation of electronic and optical concepts at the limit of single or few-atomthick materials [3]. Many current studies on 2D vdW heterostructures have been limited to the mechanically exfoliated crystals, which is experimentally arduous and an un-scalable technology for any practical application. In contrast, the controlled crystal growth and direct assembly of 2D materials is a challenging yet scalable technique, from which could emerge next-generation electronics [4].

For 2D layered transition metal dichalcogenides (TMDs) of formula $MX_2$ (M metal, X chalcogen), thickness control is particularly important as many members of the TMD family undergo a direct to indirect band gap transition from single to bilayer thickness [3]. Beyond thickness control, the actual atomic layering of the single monolayer is important, with the distinction of hexagonal (2H) and rhombohedral (3R) polytypes in TMD bilayers (2ML) [5]. The particular interest of 3R (AB-type) TMD bilayer is found in its reduced symmetry compared to 2H (AA'-type) bilayers [6], [7]. We can distinguish here two types of approach for the controlled growth of TMD crystal: the high-temperature chemical vapor deposition (CVD) technique on amorphous silica supports or high-temperature resilient substrates and the medium or low-temperature molecular beam epitaxy (MBE) approach on more varied collection of crystalline substrates. While the CVD technique has shown great success in obtaining large individual mono-crystals of comparable quality to exfoliated material [8], the control of polytypism is an experimental challenge and several bilayer polytypes often coexist on the same substrate [9]. If grown on (amorphous) silica support, the random in-plane alignment of individual crystals also results in CVD-grown material facing significant difficulties in any scalable device fabrication procedure. In contrast, the recent progresses in the direct MBE growth of TMD have shown that 2D heterostructures can be fabricated on conventional 3D semi-conductor surfaces, resulting in atomically sharp hybrid 2D/3D heterostructures [10]. This last assembly is interesting as it combines the advantages of the more established 3D semiconductors while retaining some specificities of the 2D materials. Such a hybrid approach can offer several advantages in growth control, by combining clean vdW interfaces and large-scale monocrystalline layers with a strong potential for practical industrial applications [11], [12].

Although several research works have focused on the growth and fabrication of single layer TMDs/III-V 2D/3D heterostructures [13], [14], [15], such as 1ML $WSe_2$/GaAs [10] or 1ML $MoTe_2$/GaAs [16], many possible variations have not been investigated. In particular, specific studies on hybrid 2D/3D heterostructure correlating a particular bilayer polytype are relatively uncommon. Thus, we dedicate this paper to the investigation of the electronic properties of (2D) rhombohedral-stacked bilayer $WSe_2$ on a (3D) GaP(111) substrate. Individually, layered $WSe_2$ is a 2D semiconductor with excellent optical and transport properties which has attracted great attention for future electronic devices [17], while GaP is a mature 3D III-V semiconductor with applications spanning from light emitting diodes to high-power electronics [18], [19]. Taking advantage of the monocrystalline nature of the substrate and the large-scale homogeneity of our MBE-grown layers, we show here that the coherence length of the rhombohedral domains in the plane is large enough to measure the dispersion of the electronic band structure near the valence band maximum using angle-resolved photoemission spectroscopy (ARPES). Structural characterizations based on Raman spectroscopy and atomically resolved high-resolution scanning transmission electron microscopy (HR-STEM) allows for the unambiguous determination of the majority crystal phase fraction in the grown material. The combination of ARPES and theoretical band structure calculations based on Density Functional Theory (DFT) demonstrates the synthesis of rhombohedral-stacked $WSe_2$ bilayer, over the 3D semiconductor GaP(111). Our approach represents a significant step toward the scalable synthesis of large-area rhombohedral-stacked bilayer $WSe_2$ which combines high structural quality and fine thickness control. The direct growth of such quasi-van der Waals heterostructure marks an important step toward high-performance integrated optoelectronic device and systems.

## Results and discussions:

### Crystal structure of Bilayer $WSe_2$ with AA' and AB stacking

The atomic crystal structure of $WSe_2$ consists of planes of covalently bonded W and Se atoms, with a central sub-layer of W sandwiched between two sub-layers of Se [20]. Such planar arrangement exhibits no dangling bond out-of-plane, so that multi-layer assemblies are held together through vdW forces along the out-of-plane direction [21]. In a first approximation, we consider that all individual $WSe_2$ monolayers share the same hexagonal crystal structure, but vary in their relative positioning to create several polytypes [22]. Compared to the simplest case of graphite [23], the inequivalent atoms in the $WSe_2$ sublattices allow more variation in the possible inter-layer stacking orders. Notably, favorable out of plane W/Se interactions permits $WSe_2$ to crystallize in the (2H) AA' configuration, in addition to the 3R (AB) case, the later resembling the common Bernal stacking (AB) from conventional graphite. For most TMDs, the out-of-plane interaction is so important that AA' (2H) layering is often the most stable configuration, with the (2H) AA' arrangement being the unit cell of bulk (hexagonal) $WSe_2$. For hexagonal TMD, there are actually up to five high symmetry polytypes for the two-monolayer (2ML) configuration, but 2H (AA') and 3R (AB) configurations typically show the lowest energy [5]. The atomic arrangements of $WSe_2$ AA' (2H) and AB (3R) are shown Figure 1(a). In the more symmetric AA' case, the vertical layering maximizes all the W/Se interactions along the normal direction, with the superposition of Se/W and W/Se over all the possible atomic positions. The AB configuration still make use of this stabilizing out-of-plane W/Se interaction but only the first half of the W and Se atoms are superposed with atoms from other type, the second half being isolated in the center of hexagonal rings. Consequently, to these atomic arrangements, it is possible to distinguish the (2H) and (3R) polytypes from cross section images taken along the [100] projection (Figure 1(a), bottom). There the local Se-W-Se bond orientation within each monolayer takes the form of V-shape (blue color). Depending on the respective V-shape orientation between layers, parallel or opposite, it is possible to determine the bilayer stacking order to be 3R or 2H, respectively.

For a better understanding of the $WSe_2$ 2ML band structure with 3R stacking, DFT calculations on freestanding systems are performed, see the corresponding method section for computational details, considering that GaP(111) substrate weakly modify the TMD band structure, as discussed below. Our DFT calculations confirms that the 2H polytype (AA') is the most stable bilayer configuration, followed by the 3R (AB). However, the energy difference is small and AA' only prevails by a mere 3 meV/formula unit. The corresponding electronic band structures are shown in Figure 1(b). The studied WSe$_2$ bilayers display an indirect band gap character, independently of the 2H and 3R configurations. The global valence band maxima (VBM) are situated at Γ, where the band is composed mainly by W$d_{z^2}$ and Se $p_z$ orbitals. The local VBM at K is derived from a different orbital set based on W$d_{x^2-y^2}/d_{xy}$ and Se $p_x/p_y$. Another important feature is the Q valley between Γ and K, derived from a crossover from the W $d_{x^2-y^2}/d_{xy}$ and Se $p_x/p_y$ orbitals to the W$d_{z^2}$ and Se $p_z$ orbitals, which holds the global conduction band minimum, responsible for the indirect band gap transition.

Due to different symmetries of the 3R and 2H configurations, the band structures are different. We recall that the valence band split at K for the 2H bilayer is still based on spin-orbit interaction, similar to WSe$_2$ monolayer [5]. Moreover, the two valence bands at K of the 2H polytype are spin-degenerate due to the inversion symmetry. However, in the 3R configuration, the situation is different as the W atoms from the top layer are now aligned on the top of the hollow position, while those of the lower layer still face the top Se atoms. Such absence of symmetry in the 3R configuration further splits these two bands into four spin-polarized contributions at K. This spreading of the band structure, from two bands (2H) to four bands (3R) is repeated at various positions in the band structure, with varying magnitude. The splitting at K (k=1.2 Å$^{-1}$) is actually small in energy amplitude, but stronger variations are found in the Q valley of the conduction band (k~0.7 Å$^{-1}$), Q valley in the valence band (k~0.5 Å$^{-1}$) and close to the M point in the valence band (k~1.7 Å$^{-1}$), as identified by black circles in Figure 1(b).

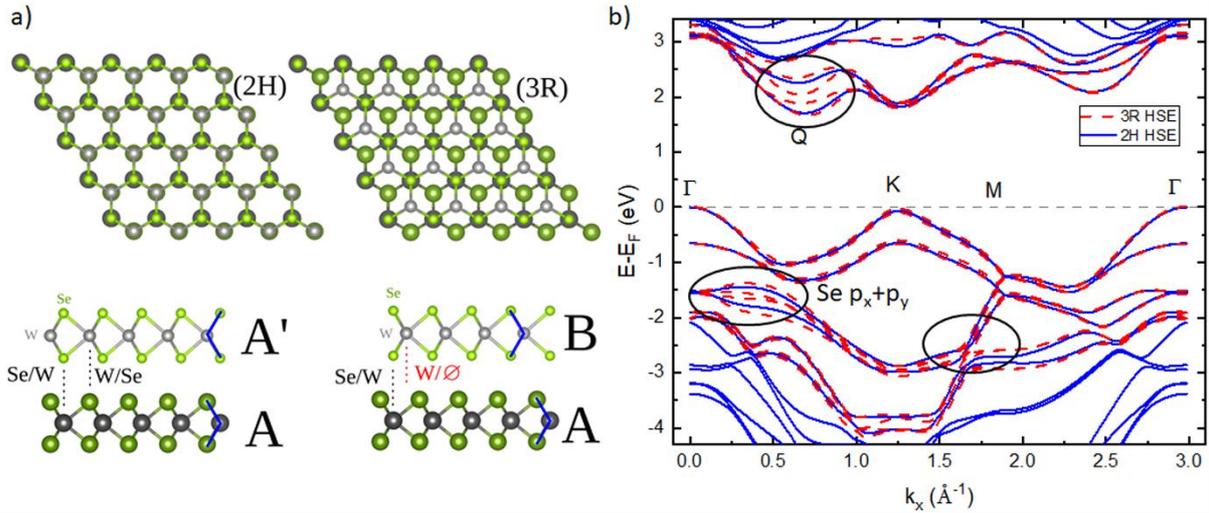

**Figure 1: Schematic representation of bilayer WSe$_2$:** Top view of (a) 2H and 3R stacking sequences of the bilayer WSe$_2$. The colored layers differentiate the successive position in the in-plane direction were each color corresponds to an inequivalent layer position, b) comparison of the band structure of 3R and 2H stacked BL WSe$_2$ using HSE calculation.

## Micro-Raman spectroscopy and STEM measurement of WSe$_2$ on Se terminated GaP(111)B

A bilayer thick WSe$_2$ film was grown on a Se terminated GaP(111)B substrate by quasi-vdW growth [10]. Such arrangement displays a quasi van der Waals character at the interface which is particularly suitable for the subsequent growth of vdW WSe$_2$ material [24]. Here, we use here commercial GaP(111)B substrates which are thermally deoxidized under P and $H_2$ flux in a dedicated MBE chamber. The oxide-free GaP(111)B substrates are then transferred in ultra-high vacuum condition to another MBE chamber where the $WSe_2$ growth is performed. Our MBE growth procedure for epitaxial $WSe_2$ growth on GaP(111) requires a Se exposure of the GaP(111)B surface at high temperature (700°C), prior to the actual $WSe_2$ growth at low temperature (250°C) using elementary W and Se solid sources. A particular characteristic of our MBE growth method is that we use low substrate temperature (250°C) to obtain the 3R $WSe_2$ stacking. The details of our complete growth procedure are listed in the methods section.

As the 2H and 3R $WSe_2$ bilayer configurations are closely related, experimental determination of the sample polytype is not straightforward. In Figure 2(a), we use a combination of both room temperature micro-Raman spectroscopy and DFT calculations to distinguish the two polytypes. Our experimental spectrum (black curve) exhibits a dominant mode at 250 cm$^{-1}$, as well as a second contribution at 246.4 cm$^{-1}$ attributed to the $E'_{2g}$ and $A_{1g}$ modes [7], and a weaker feature at 259.7 cm$^{-1}$ which is usually associated with the 2LA(M) mode [25]. To confirm the 3R character of our layer, we compare our MBE-grown $WSe_2$ (black curve) to reference CVD-grown 2H and 3R $WSe_2$ (blue and red curves, respectively) and to a 3R $WSe_2$ CVD-grown flake transferred over a graphene substrate (green curve). Using these reference CVD crystals, for which the 3R or 2H character is unambiguously determined [26], [27], we observe that $A_{1g}$ mode is dominant over the $E'_{2g}$ frequency in the in the case of 2H $WSe_2$ bilayer (blue curve). However, in the 3R configuration (red and green curves), each single $E'_{2g}$ and $A_{1g}$ mode can be resolved [25]. This Raman determination of the 3R phase is resilient and is maintained after mechanical transfer of a reference 3R crystals to a graphene host surface (green curve) [27]. The matching position of the two main lines at 250 cm$^{-1}$ and 248 cm$^{-1}$ between the reference CVD 3R sample (red curve) and our MBE $WSe_2$/GaP(111) sample (black curve) indicates that the latter shows some 3R character. We then compare the Raman peak position of our MBE-grown 3R $WSe_2$/GaP(111) sample to the nominally strain-free CVD-grown 3R $WSe_2$ mechanically transferred on graphene (green curve). We observed that the $E'_{2g}$ mode of our MBE-grown $WSe_2$/GaP(111) is shifted by about 1.4 cm$^{-1}$. Such shift of the $E'_{2g}$ in-plane Raman mode can be interpreted as small residual tensile strain about 0.6 % in the $WSe_2$ bilayer [28], [29], [30] which may originate from the epitaxial growth or from the thermal mismatch between the $WSe_2$ layer and GaP substrate. The line widths of the $A_{1g}$ and $E'_{2g}$ peaks is 2.5 cm$^{-1}$, calculated by the full width at half maximum (FWHM) after a Lorentzian fitting, which indicate that our MBE 3R $WSe_2$/GaP(111)B bilayer has a slightly larger defect density the reference CVD 3R $WSe_2$ crystals (1 cm$^{-1}$ FWHM). The relative intensities of the peaks $E'_{2g}$ and $A_{1g}$ may be also used as a measure of the 3R/2H stacking order [26]. In the current case, it suggests that our sample is typically composed of the 3R phase. Additional micro-Raman measurements performed at random locations over the sample surface do not show any difference with the selected spectrum (black curve), so that the sample can be considered homogenous at the macroscopic scale.

To confirm the stacking order of our sample, we have carried out HR-STEM, using a focused ion beam to extract a cross-sectional lamella. Figures 2(b) and (c) present HR-STEM images acquired in the High Angle Annular Dark Field (HAADF) mode to obtain chemical contrast between the $WSe_2$ and GaP materials. In the low magnification image in Figure 2(b), we quickly identify the bilayer $WSe_2$ over the GaP due to the change of HAADF contrast, relative to the large variation in average atomic number between GaP ($Z_P=15$, $Z_{Ga}=31$) and $WSe_2$ ($Z_W=74$, $Z_{Se}=34$). Over the analyzed area, (see Supplementary Information Figure S1 for a larger view) we can conclude that the $WSe_2$ thickness is regular, being exclusively 2ML thick with no region composed of 1ML or 3ML $WSe_2$. At this large scale, the $WSe_2$/GaP interface appears sharp and flat, without major defect or strong inter-diffusion between the materials. We observe a lateral coherence length of about 50 nm in WSe$_2$, indicating the presence of small domains within our WSe$_2$ bilayer. The high magnification image Figure 2(c) shows atomic resolution and is taken along the GaP [1-10] zone axis. The crystal structure of GaP is confirmed cubic and that of individual WSe$_2$ single layer as hexagonal. We note that the WSe$_2$ bilayer is fully monocrystalline and oriented on the GaP(111)B surface. We observe the following epitaxial relationship with [11-20]$_{WSe2}$//[1-10]$_{GaP}$ and [1-100]$_{WSe2}$//[11-2]$_{GaP}$ in the plane and [0001]$_{WSe2}$//[-1-1-1]$_{GaP}$ out of plane. Such orientation of WSe$_2$ on a 3D III-V substrate was also observed in the literature on the Se-terminated GaAs(111)B surface [10]. In-plane grazing incidence XRD was used to analyze the crystallographic registry between the WSe$_2$ domains and the GaP(111) substrate at large scale (probed area 1 x 0.8 cm², supplementary information figure S2). We retrieve the 6-fold XRD symmetry of the in-plane WSe$_2$ 100 and GaP 224 reflections, similarly to the work of Redwing et al [31], and we confirm the epitaxial relationship determined above with [11-20]$_{WSe2}$//[1-10]$_{GaP}$.

The interlayer spacing between the two WSe$_2$ layers is (6.45 ± 0.10) Å, which is in very good agreement with the standard vdW gap between consecutive layers in bulk hexagonal WSe$_2$ (6.48 Å) [32]. The surface of the GaP(111)B substrate is atomically sharp and the last plane of atoms shows an increased HAADF contrast, which is compatible with a Se-termination of the surface, with Se replacing P atoms ($Z_P$=15, $Z_{Se}$=34). A similar Se-passivation is reported in the literature for GaAs(111)B surface [10, 33]. Assuming the same phenomena occurs here on GaP(111)B, the Se-termination will minimize the number of dangling bonds on the initial GaP(111) surface so that the Se-terminated GaP(111)B surface presents a strong van der Waals character. The vertical separation between the central W atoms of the first WSe$_2$ layer and the lower Ga atom is (6.20 ± 0.20) Å, which is consistent with a quasi-vdW gap between WSe$_2$ and GaP(111). The in-plane lattice mismatch between WSe$_2$ ($a_{WSe2}$=3.28 Å) and the GaP substrate is evaluated to ~15% by approximating the GaP(111) surface as an hexagonal surface cell with ($a^{hex}_{GaP}$= $a_{GaP}$/√2 = 3.854 Å). In these conditions, we observe a coincidence lattice with 6 units of GaP approximately matching 7 units of WSe$_2$ (Figures 2(c-d)). This leads to a reduced average lattice mismatch (0.65%), which may be accommodated elastically. Comparing the V-shape orientation of the Se-W-Se bonds between the upper and lower WSe$_2$ layers, blue line Figures 2(c-d) see Figure 1(a) for the reference orientation, we confirm the existence of the 3R configuration in our sample, in line with the Raman investigations above.

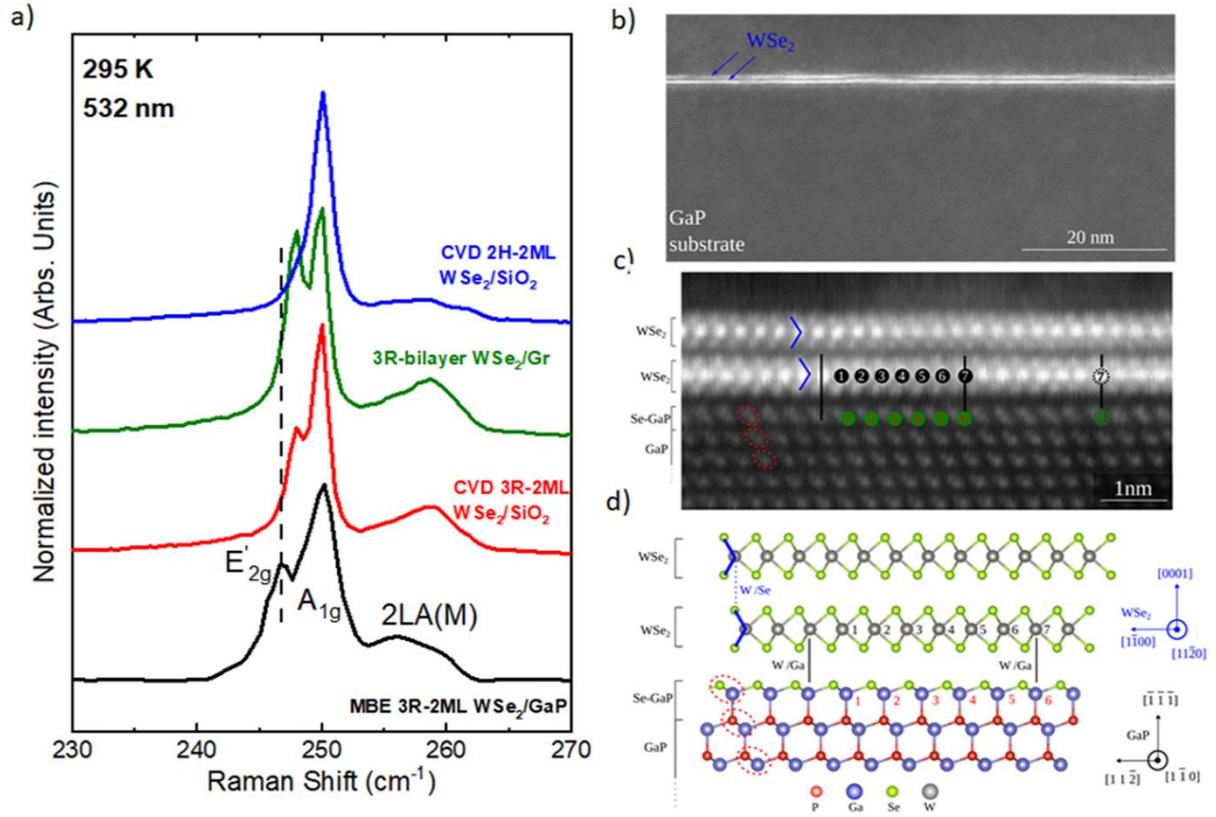

**Figure 2: Stacking order and Structural interface investigation of the WSe$_2$/GaP(111) heterostructure:** (a) micro-Raman spectroscopy of the as grown bilayer WSe$_2$/GaP(111)B (black curve), micro-Raman spectrum of reference 3R and 2H stacked bilayer WSe$_2$ obtained by CVD (red and blue curves respectively), and micro-Raman spectrum of reference CVD 3R bilayer WSe$_2$ transferred on graphene substrate (green curve). (b) Atomically resolved bright-field HR-STEM image along the (110) zone axes. (c) high resolution STEM image with the overlaid atomic structure of the bilayer WSe$_2$ on the GaP(111)B surface. (d) Schematic structure of the bilayer WSe$_2$ on the GaP(111)B surface.

## Interface properties of WSe$_2$ bilayer on Se-terminated GaP(111)B

While the previous micro-Raman and TEM data indicate that our WSe$_2$ grown on Se-terminated GaP(111) is a vdW hybrid 2D/3D heterostructure, we still need to prove the absence of covalent bonds between the different layers within our sample. To achieve this, we use X-ray photoemission spectroscopy (XPS) acquired at synchrotron facility in order to study the core levels of the different atoms [10]. For XPS measurements, the conductivity of the substrate is important to avoid the electronic charging of the film [34], [35]. Here, the growth substrate is chosen as n-type GaP(111) to be directly compatible with XPS analysis, without further treatment. XPS measurements of the bilayer WSe$_2$ reveal sharp peaks corresponding to W 4f and Se 3$d$ core levels, as shown in Figure 3(a). In addition to the expected 3R WSe$_2$ bilayer (see micro-Raman and HR-STEM above), the binding energy plot of the W 4f and Se 3d core levels points to the existence of a minor fraction of the distorted 1T' phase of WSe$_2$ [36]. Relative to XPS analysis, 3R and 1T' WSe$_2$ have different binding energy signature in the W 4f and Se 3d core levels, which allows a quantification of the 3R to 1T' fraction in WSe$_2$. We deconvolve the W 4f core levels into four components in Figure 3 (b). The W 4f main doublet (light blue curves), with the W 4f$_{7/2}$ peak at a

BE = 32.4 eV, represents the signature of WSe$_2$ 3R or 2H. The other doublets of much smaller amplitude are related to the 1T' phase [37],[36]. A similar deconvolution procedure is performed on the Se 3d core level (Figure 3 (c)) for which we observe two doublets (Se 3d$_{5/2}$ and Se 3d$_{3/2}$). The high intensity doublet Se 3d$_{5/2}$ at BE = 54.3 eV corresponds to the Se atoms embedded in the WSe$_2$ bilayer, while the Se 3d$_{5/2}$ at lower BE = 53 eV corresponds to the terminating Se atom bounded to Ga at the top surface of the GaP(111), see STEM-HAADF image and corresponding schematic figure 2(c-d). Such Se termination of the GaP(111)B surface displays a quasi van der Waals character which is particularly suitable for the subsequent growth of WSe$_2$. The other doublet of much smaller amplitude with the Se 3d$_{5/2}$ located at 53.7 eV is related to the 1T' WSe$_2$ phase [36],[37]. Using XPS data, we can't distinguish between 2H and 3R, however based on the Raman and STEM observations, we associate the main peak with 3R. The ratio of the 3R and 1T' phases was estimated from the respective peak area of the W 4f spectra from the two different phases (figure 3(b)). The XPS peak area it is proportional to the volume of the corresponding phase (given the comparable atomic densities of 3R and 1T' WSe$_2$), which indicates that over 95 % of the sample consists of WSe$_2$ in the 3R phase. We find no additional signature from other compound; i.e. no carbon or oxygen related bonds, which indicates that the surface and the heterostructure are free from contamination. The absence of other covalent bonds, confirms the quasi-vdW nature of the interlayer interaction. Combining XPS, micro-Raman and HR-STEM, we can affirm that the overwhelming fraction of the sample (>95 %) consists in a WSe$_2$ bilayer stacked in the 3R configuration. A clear epitaxial relationship of the 3R WSe$_2$ bilayer over the GaP(111) substrate is observed and we confirm the absence of covalent bonds in this hybrid 2D/3D heterostructure with individual layers all separated by quasi-vdW gaps.

**Electronic band structure of bilayer WSe$_2$**

Now that we have determined the structural properties of our sample, we can focus on the experimental determination of its electronic band structure by using ARPES [38]. The raw photoemission spectra of the valence band, measured with a photon energy of 80 eV, is shown in the insert Figure 3(a). ARPES data along the high-symmetry directions over the entire Brillouin zone (BZ) are shown in Figure 3(d). The full 3D BZ is shown in the insert with the Γ point at its center, *M* points at the midpoints of the hexagonal BZ edges, and *K* points the corners of the hexagon. In the 2D projection, $\overline{\Gamma}$, $\overline{M}$ and $\overline{K}$ are the respective projections of these points, as illustrated in the insert Figure 3(d). This image shows a clear energy cutoff at about (0.8 ± 0.1) eV above the VBM, which we identify as the position of the Fermi level (E$_F$). This confirms our 3R BL WSe$_2$ to be a semiconductor, in agreement with previous experimental reports [10],[39],[22]. The agreement between the free-standing 3R WSe$_2$ theoretical calculation and our experimental 3R WSe$_2$/Se-GaP(111) heterostructure is investigated in detail in Figure 3(e). To do so, we superpose the k–E maps of the two electronic band structures, uniformly shifting the Fermi level of the DFT data to match the experimental VBM. We observe an agreement over all the investigated k-space. The VBM is at Γ, indicating of the indirect band gap character of the bilayer WSe$_2$, in stark contrast with direct band gap monolayer WSe$_2$ [20]. The Figure 3 confirms that the 3R WSe$_2$/GaP(111) hybrid heterostructure in 3R WSe$_2$ shares the same electronic dispersion as a free-standing 3R WSe$_2$ BL. As expected from the weak hybridization in both heterostructures, the top of the valence band near K is essentially composed of states from the WSe$_2$ bilayer. The valence band also decomposes at the top of the Γ point in two well separated bands, which prove that the WSe$_2$ thickness is a bilayer over the probed area. Considering the agreement between the 3R bilayer WSe$_2$ DFT calculations and the experimental ARPES data, we confirm that the WSe$_2$ bilayer can be considered as quasi-freestanding on the GaP(111)B surface, with no interlayer hybridization.

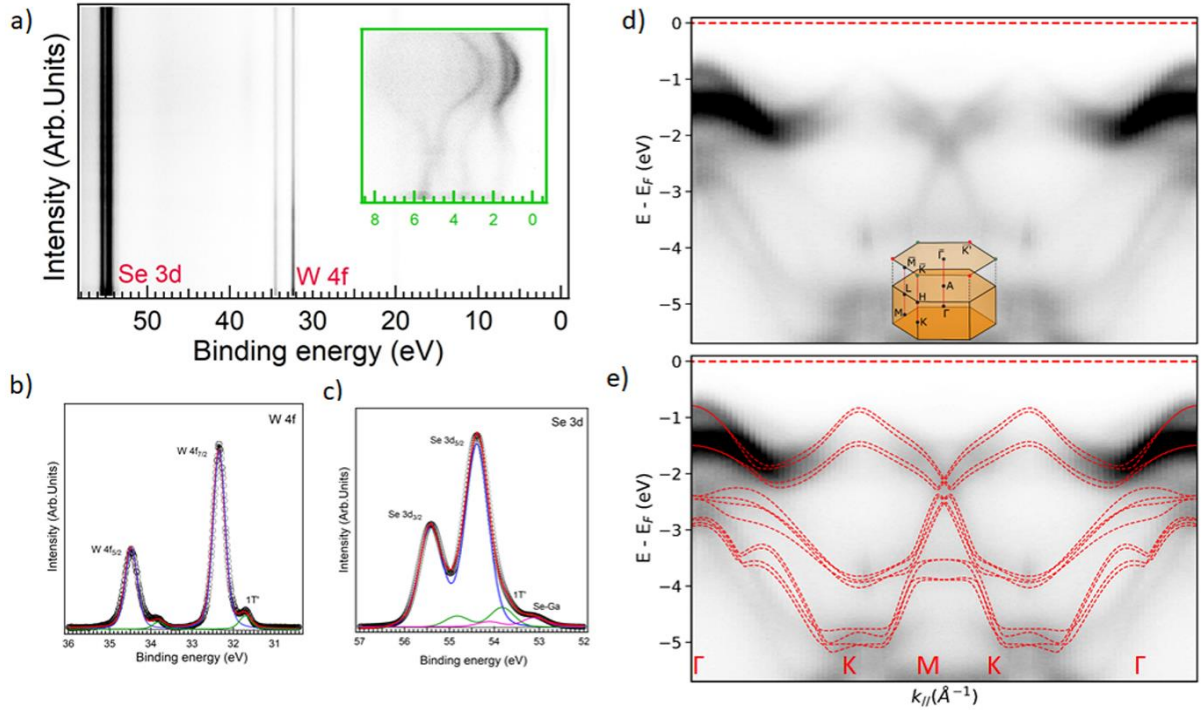

**Figure 3: Chemical properties of the WSe$_2$/GaP(111)B heterostructure:** a) XPS spectrum measured with h$v$ = 80 eV (the mapping represents the recorded intensity along the binding energy range). The inset shows the magnified valence band, b and c) XPS core level spectra of Se-3d and W-4f respectively, d and e) ARPES measurements (h$v$ = 80 eV, $T$ = 30 K) of the electronic structure along the GKMK'G high symmetry directions, e) same data with overlaid theoretical DFT calculations from free standing bilayer WSe$_2$. The Fermi level position is located at the zero of the binding energy (marked by a red dotted line).

We further explore the ARPES electronic band structure along the ΓK high-symmetry direction of the BZ using different photon energies: 50 eV, 60 eV and 80 eV. Figures 4 (a, b and c) shows the corresponding ARPES maps, respectively. The position and size of the hole pocket at Γ and K points does not change with the photon energy, and simply shows a decrease in intensity. The intensity variation between the two bands at Gamma is a signature of the possible dispersion in k$_z$. The ARPES signal at energies below -3 eV from $E_F$ highlights several dispersive bands which intensity changes with the probing photon energy. The shape and dispersion of these bands seem to be affected by modification on the probing photon energy attributed to the k$_z$ dispersion. In Figure 4(d), we show a superposition of isoenergy cuts of projected first BZ obtained in our optimized surface sensitive condition (50 eV). The particular isoenergy cut at a BE close to the VBM of WSe$_2$ reveals only electronic states at Γ and the six K and K' points. The other isoenergic cuts at higher BE indicate a circular hole pocket at the Γ point and a triangular hole pocket at each of the K and K' points. Those highest binding energy states form hole-like bands centered at the BZ corners (K), which are approximately circular pockets near the band maxima and becomes trigonally warped as they enlarge with increasing binding energy, eventually merging with the zone-center bands to form bone-shaped pockets. All K pockets arrange into a hexagonal symmetry, which indicates that they originate from the WSe$_2$ bilayer. The incident photon beam was focused into a 50−μm spot (in diameter) on the sample surface. The presence of a single pocket at the expected position of the six K/K' points, which indicates that it originates from the single orientation WSe$_2$ bilayer on the GaP substrate (Iso-energy cut at -1.3 eV in figure 4(d)). The presence of a single pocket at a unique well-determined K position in the plane further confirms that the WSe$_2$ bilayer is single orientation, with a unique lattice alignment between the WSe$_2$ and the GaP(111) substrate. Along the high symmetry ΓK direction, we resolve two split branches of the valence band of WSe$_2$ at the K point. Fitting the measured energy distribution curves (EDC) at K, we estimate the SOC is about 550 meV (Figure S3). The large experimental splitting of the valence band at K is one of the signatures of the strong atomic spin–orbit interaction in this compound, consistent with our first-principles calculations in Figure 1(b).

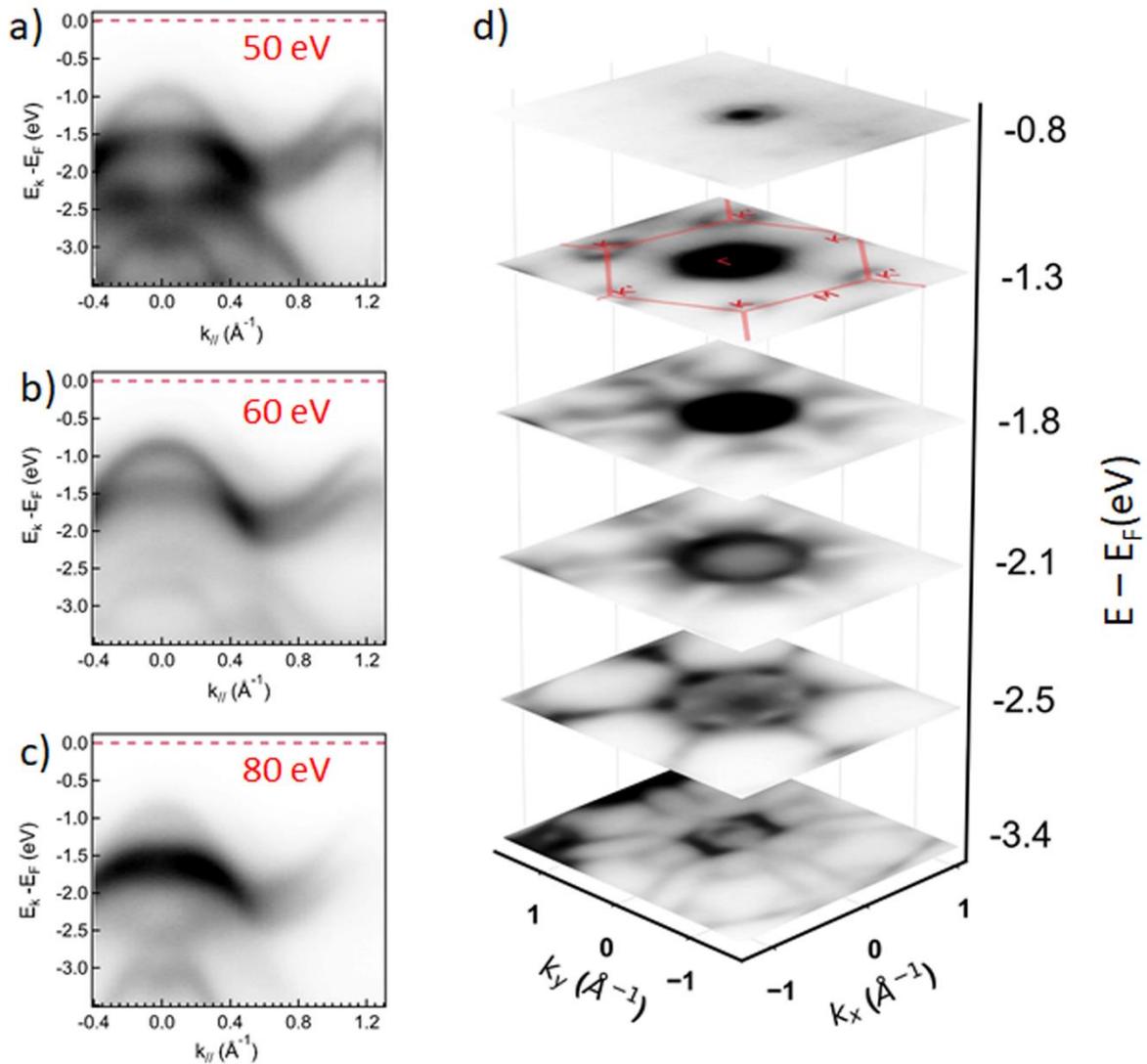

**Figure 4: ARPES measurements of the WSe$_2$/GaP(111) heterostructure:** (a, b and c) ARPES measurements of the WSe$_2$/GaP(111) heterostructure along the ΓK direction of the WSe$_2$ first Brillouin zone measured at three photon energy hν = 50, 60 and 80 eV. (d) Iso-energy cuts of the first Brillouin zone of the ΓKM plane obtained in surface-sensitive conditions (hv = 50 eV). The Fermi level position is located at the zero of the binding energy (marked by a red dotted line).

In order to confirm the particular 3R stacking order of our WSe$_2$ bilayer heterostructure, we compare our experimental ARPES data with the band structure calculations in the DFT framework for free-standing 3R and 2H WSe$_2$ bilayers. The *k–E* map of the electronic band structure of this sample along the ΓK high-symmetry direction in Figure 5(a), reveals the electronic bands of bilayer WSe$_2$. To enhance fine spectral features and obtain better clarity of the band structure presented in Figure 5(a), the second derivatives of the photoelectron intensity as a function of binding energy and parallel wave vector were also provided in Figure 5(b). The experimental data in Figure 5(b) is overlaid with DFT calculated band structures considering spin-orbit coupling for 2H (blue curve) and 3R stacking (red curve) WSe$_2$ BL in Figures 5 (c and d), respectively. As expected from previous results in Figure 3(d), all the main features are equally well reproduced by the 3R and 2H calculated band structures that we need to examine the band structure to discriminate between the two possibilities. While the DFT dispersion dispersion of the Se p$_x$+p$_y$ at -3 eV shows different number of bands between the 2H and 3R polytypes, the experimental broadening in the ARPES data makes the direct identification difficult. However, the energy separation between the two outer bands can be used as an indication, showing that the 3R stacking band structure is in slightly better agreement with the experiment than that of 2H (see insets showing the second derivative in Figure 5). This can also be better visualized by comparing the energy distribution curve (EDC), obtained by integrating the intensity map in a wavevector window of 0.05 Å−1 around k = 0.4 Å$^{-1}$, with the positions of the calculated bands of 2H and 3R Figure 5(e). Quantitatively, the value of the energy splitting is of 0.4 eV for the bilayer measured spectrum (estimated using the full width at half maximum,), whereas it is of 0.33 and 0.46 eV for the calculated bands of 2H and 3R, respectively. The energy splitting between the Se p$_x$+p$_y$ bands is thus higher for 3R than for 2H, which tends to be more consistent with the 3R configuration. The same analysis can be performed close to the M-point of the valence band (see Supplementary Information Figure S4). Therefore, the ARPES data suggests that the calculated band dispersion for 3R is in slightly better agreement with the experiment than 2H, in agreement with our structural Raman and STEM observations. To further confirm the obtainment of this 3R stacking by ARPES, a possibility not explored in this work would be to measure the unoccupied electronic band structure near the Q valley of the conduction band by means of inverse photoemission or by time-resolved ARPES. There we expect a large splitting of the Q valley between two or four bands, which may be easier to characterize experimentally.

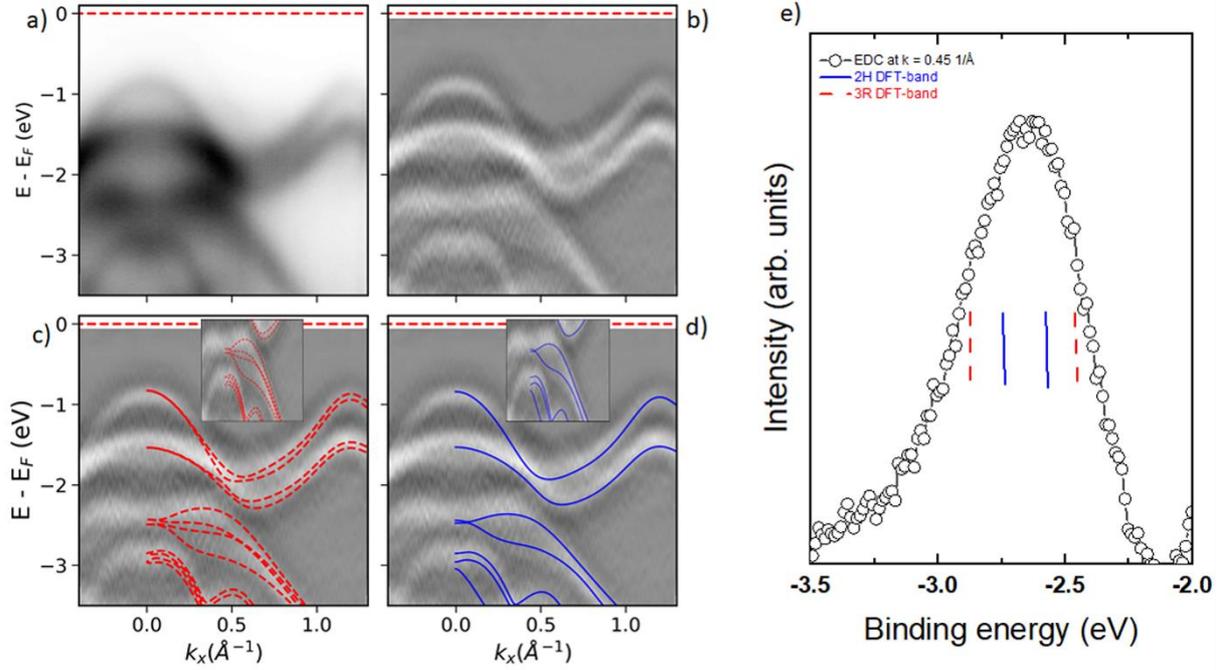

**Figure 5: Comparison between the experimental band structure of WSe$_2$/GaP(111) and DFT calculation:** a) ARPES measurements of the valence band along the ΓK high symmetry directions of WSe$_2$ respectively ($hv$ = 50, $T$ = 30 K), b) Corresponding second derivative image of the (a), (c) and (d) same data with overlaid theoretical DFT calculations from free standing 3R and 2H stacking order WSe$_2$, e) energy distribution curve (EDC), obtained by integrating the intensity map in a wavevector window of 0.05 Å$^{-1}$ at k$_{//}$ = 0.4 Å$^{-1}$.

## Conclusions

In summary, we have demonstrated the quasi van der Waals epitaxy of WSe$_2$ bilayer on GaP(111) by MBE, which results in a 2D/3D hybrid heterostructure with atomically sharp van der Waals interfaces. Combining micro-Raman, HR-STEM and XPS, we conclude that the vast majority of the sample is composed of epitaxial 3R-WSe$_2$ bilayer, with a minor fraction of 1T' WSe$_2$. We observe that the 3R-WSe$_2$ bilayer is weakly hybridized to the GaP(111) surface and that its electronics band structure is well described by free-standing 3R-WSe$_2$ calculations. Detailed analysis of the respective 2H and 3R band structure shows specific features that can allow the direct determination of the stacking order from the ARPES data. Our results show a good agreement between our experimental data and theoretical band structure of the 3R WSe$_2$ bilayer calculated from DFT using the HSE functional. While our experimental ARPES data still need support from other technique to unambiguously assign the particular 3R stacking order in our layer, this work shows that the ARPES characterization of specific dispersion feature may soon be sufficient to fully determine the polytype of a given vdW heterostructure.

## Methods:

**Experimental growth of WSe$_2$ bilayer on terminated GaP(111)B substrate:** Commercial GaP(111)B substrates (residual n-type doping 2-4.7 10$^{16}$ at.cm$^{-3}$, from ENSEMBLE3 Ltd. Poland) are cleaved into regular piece of 1/6 of 2 inch wafer to fit our various sample holders. Then each 1/6 piece GaP(111)B is deoxidized under cracked phosphine and an atomic hydrogen flux by ramping the substrate temperature up to 540°C in a dedicated III-V MBE reactor (Riber Compact 21 TM, base pressure 1.10$^{-10}$ Torr). This results in a sharp (2x2) GaP RHEED pattern, characteristic of fully-deoxidised GaP(111)B surface. The sample is then transferred under ultra-vacuum condition to another dedicated TMD reactor (Vinci Technologies, base pressure 1.10$^{-10}$ Torr) for Se passivation. The sample is heated up to 250°C before being exposed to a Se flux of 5.10$^{-6}$ Torr (Riber VCOR 110 valve cracker cell) and further annealed to 700°C and kept at this temperature for 3 minutes. This leads to a sharp (1x1) RHEED pattern and allows obtaining a Se-termination of the GaP(111)B top surface as checked by XPS. Such arrangement displays a quasi van der Waals character which is particularly suitable for the subsequent growth of lamellar vdW WSe$_2$ material. The WSe$_2$ growth is calibrated to obtain a final layer of 2ML thickness. For this, the temperature of the Se-terminated GaP(111)B surface is ramped down and stabilized at 250°C under the Se flux before being exposed to the W flux from an electron gun evaporator (Telemark model 575). The W growth rate of 1.8 10$^{-3}$ Å/s is calibrated by a quartz crystal monitor. The RHEED pattern remains streaky all along the growth duration at 250°C. The samples are capped with amorphous Se to protect them from surface oxidation. This capping was then removed thermally prior to the XPS and ARPES measurements.

**Raman spectroscopy:** Raman spectroscopy measurements were performed on a Horiba Scientific LabRAM HR at an excitation of $\lambda$ = 532 nm, in a backscattering geometry in parallel-polarized configuration, with a 360° rotational sample stage. The spectral resolution is $\sim$ 0.7 cm$^{-1}$ for the grating of 1800 grooves per mm. The spectrometer was calibrated with a pristine silicon sample.

**Photoemission spectroscopy.** ARPES experiments were performed at the CASSIOPEE beamline of the SOLEIL synchrotron light source. The CASSIOPEE beamline is equipped with a Scienta R4000 hemispherical electron analyzer whose angular acceptance is ±15° (Scienta Wide Angle Lens). The experiment was performed at T = 30 K. The total angle and energy resolutions were 0.25° and 16 meV, respectively [40].

**Electronic structure calculation:** VASP package [41], [42], [43] was used within the projector augmented-wave formalism [44], [45] with fourteen and six electrons explicitly included in the valence states. for W and Se atoms respectively. The *k*-points sampling of the first Brillouin zone was done with a **Γ**-centered grid of (12 × 12 × 1) points, with a small gaussian smearing (0.05 eV) to handle partial occupancies of the bands. Optimized geometries have obtained using PBE-D3 functional [46], to treat carefully weak van der Waals forces between the two layers. The lattice parameter was fixed to 3.32 Å with sufficient vacuum height (~22 Å), to avoid spurious interaction between periodic images in *z*-direction. On top of those calculations, Heyd-Scuseria-Ernzerhof (HSE06) hybrid functional [47], [48], [49] calculations were performed, including spin-orbit coupling effects. WANNIER90 program [50] was used to yield band structures within Wannier interpolation framework.

**ACKNOWLEDGMENTS**

The authors thank L. Largeau and N. Findling at C2N for their timely help during the X-ray diffraction analysis. We acknowledge the financial support by MagicValley (ANR-18-CE24-0007), Graskop (ANR-19-CE09-0026), 2D-on-Demand (ANR-20-CE09-0026), MixDferro (ANR-21-CE09-0029), Tunne2D (ANR-21-CE24-0030) grants and (ANR-22-PEXD-0006) FastNano project, as well as the French technological network RENATECH. D.R. acknowledges support from the HPC resources of IDRIS, CINES and TGCC under the allocation 2022-A0120912417 made by GENCI. I. C. G. thanks the CALMIP initiative for the generous allocation of computational time, through Project No. p0812, as well as GENCI-IDRIS and GENCI-TGCC, Grant No. A012096649.

**Supporting Information:** Large view STEM-HAADF of the sample. In plane XRD performed on the GaP 112 and WSe$_2$ 100 reflections to verify the epitaxial relationship. Energy curve analysis (EDC) of the ARPES spectra as reported Figure 5. Comparison between ARPES experimental and theoretical band structure of 3R and 2H bilayer WSe$_2$ is also shown along the KMK' direction.

**Competing financial interests:** There are no conflicts to declare.

# Supplementary information

## Quasi van der Waals Epitaxy of Rhombohedral-stacked Bilayer WSe$_2$ on GaP(111) Heterostructure

Aymen Mahmoudi[1], Meryem Bouaziz[1], Niels Chapuis[2], Geoffroy Kremer[1], Julien Chaste[1], Davide Romanin[1], Marco Pala[1], François Bertran[3], Patrick Le Fèvre[3], Iann C. Gerber[4], Gilles Patriarche[1], Fabrice Oehler[1], Xavier Wallart[2], Abdelkarim Ouerghi[1*]

[1]Université Paris-Saclay, CNRS, Centre de Nanosciences et de Nanotechnologies, 91120, Palaiseau, Paris, France
[2] Univ. Lille, CNRS, Centrale Lille, JUNIA ISEN, Univ. Polytechnique Hauts de France, UMR 8520-IEMN F59000 Lille France
[3]Synchrotron SOLEIL, L'Orme des Merisiers, Départementale 128, 91190 Saint-Aubin, France
[4]Université de Toulouse, INSA-CNRS-UPS, LPCNO, 135 Avenue de Rangueil, 31077 Toulouse, France

Corresponding Author: abdelkarim.ouerghi@c2n.upsaclay.fr

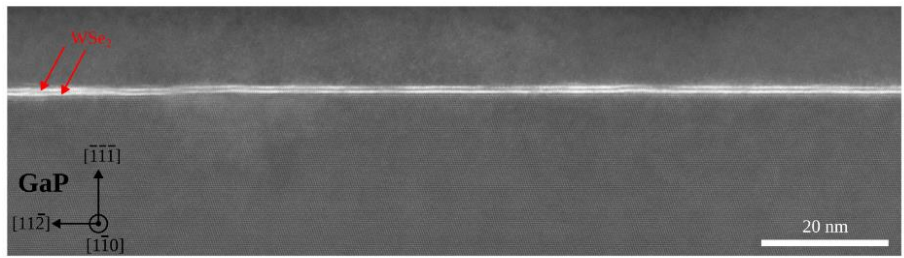

**Figure S1:** Large-scale image of an atomically resolved bright-field HR-STEM image taken along the [110]-zone axis of GaP(111)B.

In-plane XRD was performed on the WSe$_2$ 100 (2θχ = 31.22°) and GaP 224 (2θχ=87.22°) reflections using a Rigaku Smartlab equipped with a rotating Cu anode. The normalized counts have been offset for clarity. The absolute in-plane angle φ$_{abs}$ is computed from the experimental φ parameter with φ$_{abs}$ = φ -2θχ /2. We retrieve the 6-fold XRD symmetry of the in-plane WSe$_2$ 100 and GaP 224 reflections. A single set of 6 reflections is observed for WSe$_2$ 100, in agreement with the single epitaxial relationship found by TEM.

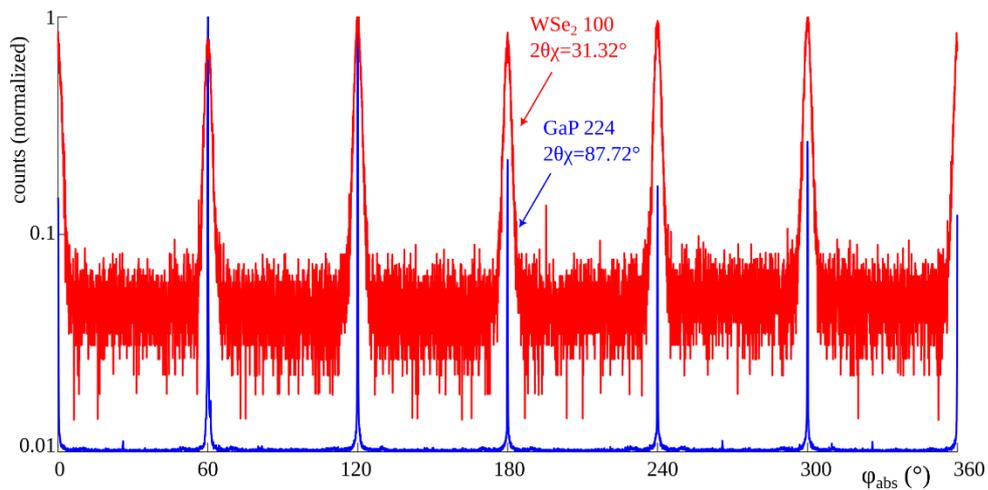

**Figure S2:** Epitaxial relationship determined by in-plane XRD: Normalized in-plane XRD showing φ -scan of the GaP 224 (2θχ = 87.22°) and WSe$_2$ 100 (2θχ = 31.32°) reflections against the in-plane angle φ$_{abs}$.

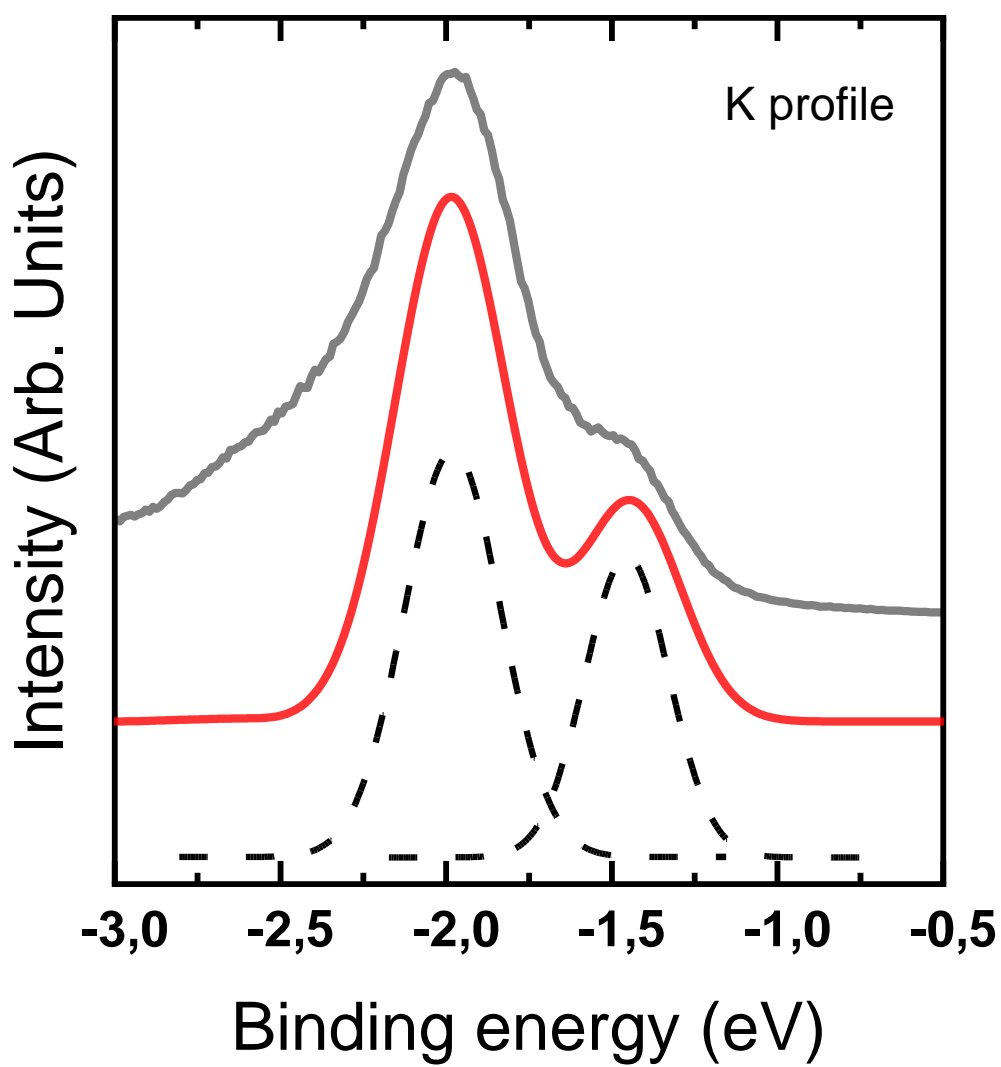

**Figure S3:** Energy distribution curve (EDC), obtained by integrating the intensity map in a wavevector window of 0.05 Å$^{-1}$ near the K point.

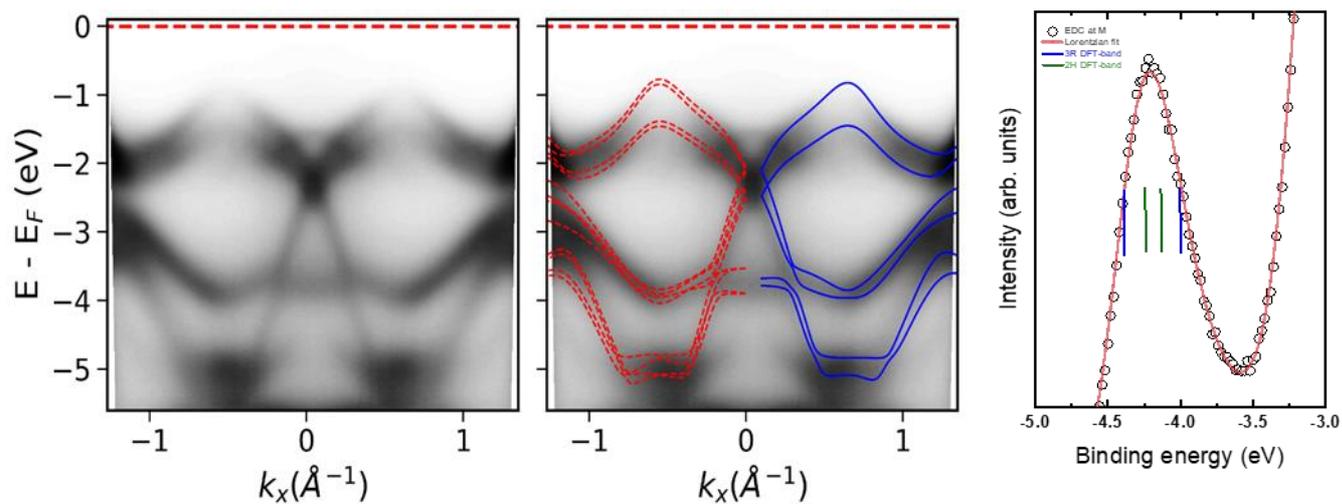

**Figure S4:** (a) Experimental ARPES measurement of bilayer WSe$_2$/GaP(111) along KMK', theoretical dispersion of the 3R and 3R (red) and 2H (blue) are superposed on the right panel. (b) Energy distribution curve (EDC), obtained by integrating the intensity map in a wavevector window of 0.05 Å$^{-1}$ around the M point.